\documentstyle[aps,preprint,prl]{revtex}

\begin{document}   
\draft
\preprint{ }
\title{Manifestations of the pseudogap in 
the BOSON-FERMION model for Bose-Einstein condensation driven 
Superconductivity}
\author{J. Ranninger and J. M. Robin }
\address{Centre de Recherches sur les Tr\`es Basses Temp\'eratures, 
Laboratoire Associ\'e \`a l'universit\'e Joseph Fourier, Centre 
National de la Recherche Scientifique, BP 166, 38042 Grenoble 
C\'edex 9, France}
\maketitle

\vspace{0.5cm}

\centerline{ ( \today )}

\begin{abstract}
The normal state behaviour of the density of states of the 
electrons  described by the BOSON - FERMION model for 
Bose-Einstein condensation driven  superconductivity 
is characterized by the 
appearence of a pseudogap which developes into a true gap
upon lowering the temperature and the superconducting critical 
temperature is approached. The consequences of this on the 
temperature dependence of the specific heat, the NMR relaxation 
rate and the optical conductivity is 
examined.

\end{abstract}

\vspace{2cm}


\narrowtext

The opening of a pseudogap in the density of states (DOS) of the 
electrons
in the normal state of high $T_c$ superconductors (HTcSC) is one of 
the characteristic features 
of these materials\cite{Battlogg-94}. 
One possible interpretation involve spin fluctuations as the 
underlying mechanism. This pseudogap is then called the spingap,
the discussion of which has given 
rise to intense work based on the $t-J$ model or the nearly antiferromagnetic
Fermi liquid.
In this Letter, we want to adress an alternative mechanism for this
pseudogap in terms of superconducting fluctuations.
The existence of such a pseudogap  together with the experimental 
indications that possibly 
two types of charge carriers (fairly localized ones and itinerant ones) 
are involved in the high $T_c$ phenomenon\cite{Timusk-93} supports a 
scenario of a mixture of intrinsically localized Bosons 
(tightly bound electron pairs) and itinerant electrons (Fermions).
This situation can be  described in its simplest form by the so called  
BOSON - FERMION (BM) model which 
was first introduced in connection with the 
many polaron problem in the cross-over regime between adiabatic 
and non-adiabatic behaviour. 
In such a scenario\cite{Ranninger-PhysicaB-85} bipolarons (Bosons) 
are envisaged to coexist with quasifree electrons (Fermions) 
and an exchange coupling 
between the Bosons and the Fermions is assumed by which Bosons 
can decay into pairs of itinerant Fermions and vice versa. 
The physically interesting regime of parameters of this model 
is that where the superconducting state below a certain critical 
temperature $T_c$ is controlled by a condensation of the 
Bosons\cite{Ranninger PhysicaC-95} and thus can in principle lead 
to rather high values of $T_c$. This happens when the Boson level 
lies close to the Fermi level of the Fermionic subsystem. 
Assuming the exchange coupling to be local, 
we have shown\cite{Ranninger-PRL-95} how, upon lowering the temperature, 
a pseudogap in the DOS of the Fermions gradually opens up 
- ultimately developing into a true gap below $T_c$. 
The opening of such a pseudogap is driven by the onset of itinerancy 
of the intrinsically bare localized Bosons 
due to a precursor effect of their superfluidity which implies 
a concomitant onset of strong local pairing of the itinerant Fermions. 
The increased correlations of the Fermions into Fermion-pairs 
results in a DOS for single particle excitations which, close 
to the Fermi level, is drastically 
diminished and thus leads to the appearance of a pseudogap and 
single particle excitations which show  strong deviations 
from Fermi liquid behaviour\cite{Ranninger-PRL-95}.
The underlying BF model on which this behaviour has been studied 
sofar is given by the following Hamiltonian

\[
H \; = \; (zt - \mu) \sum_{i,\sigma} c_{i\sigma}^{+} c_{i\sigma} \; - \;
t \sum_{<i\neq j>, \sigma} c_{i\sigma}^{+} c_{j\sigma} \; + \; (\Delta_{_B}
- 2 \mu) \sum_{i} b_{i}^{+} b_{i} 
\]
\begin{equation}
\; + \; v \sum_{i} [ \; b_{i}^{+} c_{i\downarrow} c_{i\uparrow} \; + \; 
c_{i\uparrow}^{+} c_{i \downarrow}^{+} b_{i} \; ]
\label{Equ1}
\end{equation}
where  $c^{(+)}_{i,\sigma}$ and $b^{(+)}_i$ refer to the Fermion 
and Boson annihilation (creation) operators at site $i$ and $\sigma$ 
denotes the spin quantum number. t represents the bare 
hopping integral for tight binding electrons, 
$\Delta_B$ the energy level for the bare localized Bosons 
and $v$ the local Boson-
Fermion pair exchange. 
The chemical potential $\mu$ is taken to be common to both the
Bosons and Fermions such as to ensure charge conservation 
during the Boson-Fermion exchange. 
We have previously evaluated the single particle
Boson and Fermion spectral properties for a 1D system within the lowest order 
fully selfconsistent conserving approximation\cite{Baym-PR-62} and have shown 
explicitely the opening of the pseudogap and the destruction of 
Fermi liquid properties\cite{Ranninger-PRL-95}. 
In order to ascertain that these features were indeed unrelated to any physics 
of one dimensional systems we further 
considered the case of a 2D square 
lattice\cite{Ranninger-SSC-96} and obtained results 
which are qualitatively analogous the the 1D case thus confirming 
that the pseudogap in the DOS of the Fermions is an intrinsic feature 
of the BF model.

It is the purpose of this Letter to demonsrate how this opening of the 
pseudogap in the DOS of the Fermions and 
the destruction of the Fermi liquid properties show up
in physically accessible thermodynamic 
(specific heat, compressibility), magnetic (NMR relaxation rate, 
spin susceptibility) and transport (optical conductivity) properties.
We shall use the same approximative scheme as that which served us 
for the evaluation of the single particle 
properties previously\cite{Ranninger-PRL-95,Ranninger-SSC-96} 
and which is based on a fully selfconsistent lowest order evaluation 
of the thermodynamic potential, given by the closed loop diagram 
illustrated in Fig.1 and presenting a functional of the full one particle
Fermion et Boson Green's functions.
 Within such a scheme the one and two particle 
Green's functions are derived from this closed loop diagram 
by standard functional derivatives with respect to an external 
space-time varying field\cite{Baym-PR-62} which for our 
approximation gives rise to the following expressions for the 
Boson and Fermion selfenergies:
\[
\Sigma_{F}({\bf k}, \omega_{n}) \; = \; - \frac{v^{2}}{N} \; \sum_{{\bf q}, 
\omega_{m}} \; G_{F}({\bf q}-{\bf k}, \omega_{m} - \omega_{n}) \;
G_{B}({\bf q}, \omega_{m})
\]
\begin{equation}
\Sigma_{B}({\bf q}, \omega_{m}) \; = \; \frac{v^{2}}{N} \; \sum_{{\bf k}, 
\omega_{n}} \; G_{F}({\bf q}-{\bf k}, \omega_{m}-\omega_{n}) 
\; G_{F}(
{\bf k}, \omega_{n})
\label{Equ2}
\end{equation}
where $G_{B}({\bf q}, \omega_{m}) \,=\, [ i \omega_{m} \, - \, E_{0} \, - 
\Sigma_{B}({\bf q}, \omega_{m})\,]^{-1}$ 
and $G_{F}({\bf k}, \omega_{n}) \, = \,[ i \omega_{n} \, - \, 
\epsilon_{{\bf k}} \, - \, \Sigma_{F}({\bf k}, \omega_{n}) \,]^{-1}$ 
represent the fully selfconsistently determined Fermion and 
Boson one particle Green's functions. 
The selfconsistent set of Eqs.(2) are solved numerically for 
a square lattice with sizes up to  41$\times$41 and for a set of Matsubara
frequencies $\omega_{n}$ with $n$ up to $100$. 
This turns out to be enough to cover 
a wide enough temperature regime in order to track the evolution 
of the pseudogap in the DOS and its repercussions on the physical 
quantities which we want to discuss here. 
For computational reasons we work as usual with the difference between
the total Green's function and its zero order approximation;
i.e for $v=0$\cite{trick}.
In order to treat the physically most interesting situation 
of the BF model (where the superconducting phase is essentially due 
to a Bose condensation of the Bosons) we choose the model parameters such 
that the Bosonic level lies well inside the Fermion band and the number 
of Bosons per site 
$n_B=\sum_{i}\langle b^{\dag}_{i}b^{\phantom{dag}}_{i}\rangle$
is comparable to the number of Fermions per site 
$n_F= \sum_{i,\sigma} 
\langle c^{\dag}_{i\sigma}c^{\phantom{dag}}_{i\sigma}\rangle$.
For that purpose we choose  as representative parameters for our 
numerical work: $\Delta_B=0.4$, $v=0.1$ in units of 
the bandwidth $8t$ and $n_{tot}= 2n_B\,+\,n_F\,=\,1$ per site.

The properties of the one particle spectral functions for the 
Bosons and Fermions 
have adequately been dealt with previously and we refer the reader 
to refs[5,7]. We hence shall not discuss them here in any 
further detail but rather concentrate on the evaluation of the 
specific heat, the NMR relaxation rate and the optical 
conductivity and show to what extent they are influenced by the 
opening of the pseudogap in the DOS and the 
breakdown of Fermi liquid properties of the Fermions. We evaluate 
for that purpose the total free energy $F\,=\,E-\mu N_{tot}-TS$ where 
the inner energy 
$E\,=\,\langle H_0 \rangle + \langle vH_1 \rangle +\mu N_{tot}$
is separated into a component of the uncoupled BF system and 
into that of Boson-Fermion exchange coupling. $N_{tot} \, =\,n_{tot} N$
where N is the total number of sites in the system. The expression for 
the exchange coupling contribution to $F$ can be obtained directely by
evaluating the closed loop diagram (Fig.1) which yields
\begin{equation}
\langle vH_1 \rangle \,=\, -\frac{2}{\beta} \sum_{{\bf q}, 
\omega_{m}} \Sigma_B({\bf q},\omega_{m}) G_B({\bf q},\omega_{m})
\end{equation}
Inserting the solutions of Eq.2 into the above expression we evaluate 
F using 
\begin{equation}
F \; = \; F_{0} + \int_{0}^{v} \frac{d\lambda}{\lambda} \langle \lambda
H_{1} \rangle 
\end{equation}
where $F_{0}$ is the free energy of the non - interacting system,
and consecutively derive the specific heat at constant volume 
$C_V\,=\,(dE/dT)$ 
and the entropy $S \,=\, (E-\mu N_{tot} - F)/T$\cite{check}.
As can be seen from the temperature dependence of 
the DOS of the Fermions at the Fermi level $N(0)$(see Fig.2), 
a pseudogap starts opening up below a certain characteristic 
temperature $T^{\star}$ which for our choice of parameters is around 0.015. 
$T^{\star}$ shows up noticeably in the temperature behaviour of $C_V$ 
where it corresponds to a net upturn of  $C_V$ which, with lowering the 
temperature below $T^{\star}$, increases as $ln T$. This behaviour can 
be traced back to the onset of a precursor to superfluidity of the 
Bosons which aquire coherency i.e. quasi free particle 
like behaviour with an effective mass which diminishes as the 
superconducting state is approached\cite{Ranninger-PRL-95,Ranninger-SSC-96}.
$T^{\star}$ is equally visible in the temperature behaviour of the entropy $S$
which at this temperature shows a noticeable deviation from linearity which is 
observed for higher temperatures and is due to effectively free Fermions.
Both the inverse Boson mass and the compressibility show a 
monotonic increases with decreasing temperature
with a cross-over to a much steeper rise below $T^{\star}$. The lowest
temperature results for the specific heat, as well as for the Boson 
correlation function  show a critical behaviour with a finite value
of $T_{_C}$. We identify this transition, as it should be, as a 
Kosterlitz-Thouless transition\cite{Kost-Thoul}, since one expects
a Bose-Einstein condensation for a $2D$ system\cite{Fisher}.

The onset of a pseudogap and a concommitant coherence of the Bosons 
is also visible in the magnetic response of the system measured by 
the magnetic susceptibility
\begin{equation}
\chi ({\bf q},\omega) \,=\, \frac{1}{2\pi\,i\,\hbar}
\int d\tau e^{i\hbar\omega\tau} \;
\Theta(\tau)\langle [S^-({\bf q},\tau),S^+({\bf -q},0)] \rangle
\end{equation}
where $S^+({\bf q},\tau)\,=\,c^+_{{\bf q}\uparrow}(\tau) 
c^{\phantom{+}}_{{\bf q}\downarrow}(\tau)$ 
and $S^-({\bf q},\tau)\,=\,(S^+({\bf q},\tau))^{\dag}$.
Due to the Boson-Fermion exchange coupling local magnetic correlations are
induced among the bare uncorrelated electrons arising from the singlet 
character of the Bosons. The onset of the long range superfluid
coherence of the Bosons  leads to an onset of long range magnetic 
correlations which can be seen in the static homogeneous susceptibility 
$S_O\,=\,\frac{1}{2\pi}\chi(0,0)$ and the NMR relaxation rate 
$\frac{1}{T_1}\,=\,\frac{k_BT}{2\pi} 
\sum_{\bf q} \chi"({\bf q},\omega)/\omega$
where $\chi"({\bf q},\omega)\,=\, \ Im\chi({\bf q},\omega)$. Evaluating
the expression Eq(4) to within lowest order i.e,  
neglecting vertex corrections but fully taking into account the 
selfconsistent expressions for the Fermion one particle Green's function
we obtain the results for $(T_1T)^{-1}$ as a function of temperature
as illustrated in Fig.(3). We again notice a drastic changeover of a 
fairly  well represented temperature independent Koringa behaviour for 
$T>T^{\star}$ to a rapid drop of $1/(T_1T)$ below $T^{\star}$. Nevertheless 
even for $T>T^{\star}$ the usual Korringa ratio  $(T_1T)^{-1}/S^2_0$ does 
turn out not to be temperature independent as it should be expected 
for free uncorrelated Fermions. 
On the contrary $(T_1T)^{-1}/S_0$ is fairly temperature independent 
for $T>T^{\star}$ as can be seen from Fig.(3) and tracks 
the temperature behaviour of $N(0)$.

As the last manifestation of the pseusogap in the DOS of the Fermions
we want to discuss here the optical conductivity which is defined by
\begin{equation}
\sigma^{\alpha \beta}(\omega)\,=\,\ Im\frac{1}{i\hbar\omega}
\int \frac{d\tau}{2 \pi} e^{i \hbar \omega \tau}\Theta(\tau)
\langle j^{\alpha}(\tau) j^{\beta}(0)\rangle
\end{equation}
where the $\alpha's$ component of the total current is given by
$j^{\alpha}(\tau)\,=\,i\frac{et}{\hbar}\sum_{i,\delta}\delta^{\alpha}
c^+_{i+\delta \sigma}(\tau)c^{\phantom {+}}_{i\sigma}(\tau)$. 
e denotes the charge of the Fermions and $\delta^{\alpha}$ 
the $\alpha's$ component of the lattice vectors 
linking nearest neighbor sites. Evaluating
the isotropic optical conductivity 
$\sigma(\omega)=\frac{1}{3}\sum_{\alpha}\sigma^{\alpha \alpha}(\omega)$ 
within the lowest order approximation
(neglecting vertex corrections but fully taking into account the 
selfconsistently determined Fermion one particle Green'functions) we 
obtain the optical conductivity as a function of frequency which 
for different temperatures is plotted in Fig.(4). In the inset of 
Fig.(4) we plot the dc conductivity for different temperatures and 
notice that upon decreasing the temperature one passes at $T^{\star}$
from a metallic like behaviour to one which has activated 
semiconducting like behaviour as a result of the opening of the 
pseudogap below $T^{\star}$.
These features are also present in the optical conductivity 
which for temperatures below $T^{\star}$ shows a shift of the 
oscillator strength from the frequency regime 
$\omega \leq \omega^{\star} \simeq 2T^{\star}$ to 
$\omega \geq \omega^{\star}$ for $T\leq T^{\star}$, while for 
$T\geq T^{\star}$ a similar shift is observed in the opposite direction.
The emptying out of the spectral weight of $\sigma(\omega)$ for 
$ \omega \leq \omega^{\star}$ which would show up as a dip in the optical 
conductivity can only be approached without being reached 
because of computational difficulties in reaching 
sufficiently low temperatures.

These manifestations of the pseudogap of the DOS of the Fermions 
and in particular those seen in the magnetic\cite{NMR} 
and transport\cite{transport} response functions 
are very reminiscent of what is observed in the normal state 
of underdoped $HT_cSC$.
Such features have been previously attempted to be interpreted in the 
framework of the negative $U$ Hubbard model\cite{U<01,U<02}. 
While both models lead to pseudogaps in the one particle spectrum 
and give similar results as far as the magnetic 
response is concerned, the physics of those two models is 
nevertheless quite different. In the $U<0$ Hubbard model electron pair
states exist as "fluctuating states" with very short life times for long
wavelength excitations and the Fermiliquid properties of the 
one particle excitations are conserved\cite{U<02}. The increase in the 
value of $T_c$, obtained in the intermidiary coupling regime, is then
due to a strengthening of the Cooperpair correlations rather than to a 
Bose condensation of electron pairs. In the BF model on the contrary
electron  pairs with total momentum close to zero are longlived, condense
upon lowering the temperature and give rise to  deviations 
from Landau Fermi liquid properties. Those effects on the 
thermodynamic and transport
properties have been studied here and should be noticably 
different from those obtained on the basis of the $U<0$ Hubbard model.

\begin{figure}
\caption{Diagram for thermodynamic potentiel. Solid lines denote the
propagators for renormalized Fermion Green's functions and the wavy line 
for the fully renormalized Boson Green's function.  }
\label{Fig1}
\end{figure}

\begin{figure}
\caption{Specific heat $C_{V}$, entropy $S$ 
and density of states at the Fermi
level $N(0)$ as a function of temperature. Entropy and specific heat are
normalized to their values at $T=0.1$ and the density of states is 
normalized to it's value at $T=0.02$. }
\label{Fig2}
\end{figure}

\begin{figure}
\caption{NMR relaxation rate $1/ T_{1} T$ and the ratio $1 / T_{1} T S_{0}$
(in arbitrary units) as a function of temperature.  }
\label{Fig3}
\end{figure}

\begin{figure}
\caption{Optical conducitivy (in arbitrary units) 
$\sigma(\omega)$ as a function of
frequency (in units of the bandwith $8t$). Also shown in the inset is
the value at $\omega=0$ as a function of temperature. }
\label{Fig4}
\end{figure}

\end{document}